\input{aipcheck}

\documentclass[
    ,final            
  ]
  {aipproc}

\layoutstyle{8x11single}

\input{psfig.sty}
\begin{document}
\title{\vspace{-2em}Diffractive and Total Cross Sections at Tevatron and LHC
{\small{Presented at Hadron Collider Physics Symposium 2006, May 22-26, Duke University, Durham NC, USA}}}


\author{Konstantin Goulianos}{
  address={The Rockefeller University, 1230 York Avenue, New York, NY 10021, USA}}
\begin{abstract}
 Results from the Fermilab Tevatron $\bar pp$ collider on diffractive and total cross sections are reviewed with emphasis on physics significance and properties pointing to expectations at the LHC. 

\end{abstract}

\maketitle
\vspace*{-2em}
\section{Introduction}
\begin{figure}[h]
\psfig{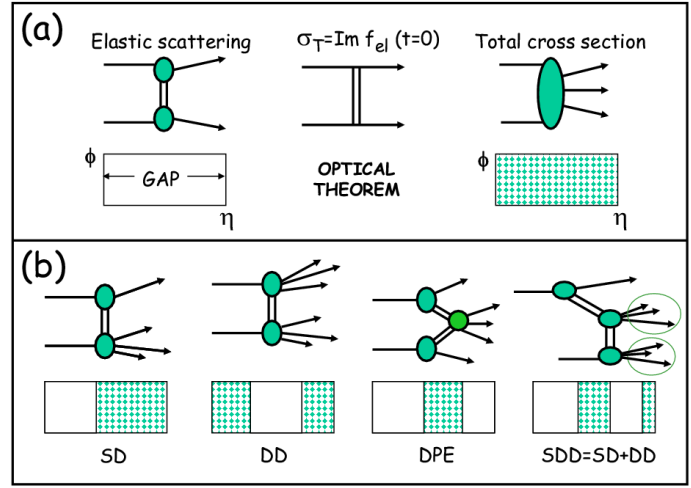}
\caption{Schematic diagrams and event topologies in azimuthal angle $\phi$ versus pseudorapidity $\eta$ for (a) elastic and total cross sections, and (b) single diffraction (SD), double diffraction (DD), double Pomeron exchange (DPE), and double plus single diffraction cross sections (SDD=SD+DD). The hatched areas represent regions in which there is particle production.}
\label{fig:diagrams}
\end{figure}

Measurements of total and diffractive hadronic cross sections at high energies provide an arena on which fundamental concepts of quantum mechanics and theoretical developments on non-perturbative QCD can be tested by confronting experimental results. In this paper, we present results obtained at the Tevatron and comment on their physics significance. Trends observed in the data are used to make predictions for expectations at the Large Hadron Collider (LHC). The presentation is mainly based on a comprehensive set of results obtained by the Collider Detector at Fermilab (CDF) since the beginning of Tevatron operations in 1989 from studies of the peocesses shown in Fig.~\ref{fig:diagrams}.   

This paper is organized in three sections: total cross sections, diffraction, and exclusive production. Recent results from CDF on the $x$-Bjorken and $Q^2$ dependence of the diffractive structure function, and on the $t$-dependence of diffractive cross sections, address questions on the QCD nature of the Pomeron; and results on exclusive di-jet and di-photon production are used to calibrate predictions for exclusive Higgs boson production at the LHC. 
\section{Total Cross Sections}
Measurements of the total cross section and $\rho$-value (ratio of real to imaginary part of the forward elastic scattering amplitude) at high energies can provide valuable input in testing fundamental concepts of quantum mechanics. 
\vspace*{-0.25em}
\paragraph {Physics issues}
\vspace*{-0.35em}
\begin{itemize}
\addtolength{\itemsep}{-0.25em}
\item Froissart unitarity bound: $\sigma_T<C\cdot\ln^2s$
\item Optical theorem: $\sigma_T\sim Im\,f_{el}(t=0)$
\item Dispersion relations: $Re\,f_{el}(t=0)\sim Im\,f_{el}(t=0)$
\end{itemize}
\vspace*{-0.25em}
The combined measurement of $\sigma_T$ and $\rho$-value provides information on $Re\,f_{el}(t=0)$, which can be used to test dispersion relations. A disagreement with theory could be interpreted as evidence for new physics~\cite{ref:Khuri}. Thus, measurements of $\sigma_T$ and $\rho$ at the LHC could open a new  window to physics beyond the standard model. 

\vspace*{6em}
{\hspace*{-6em}\psfig{figure=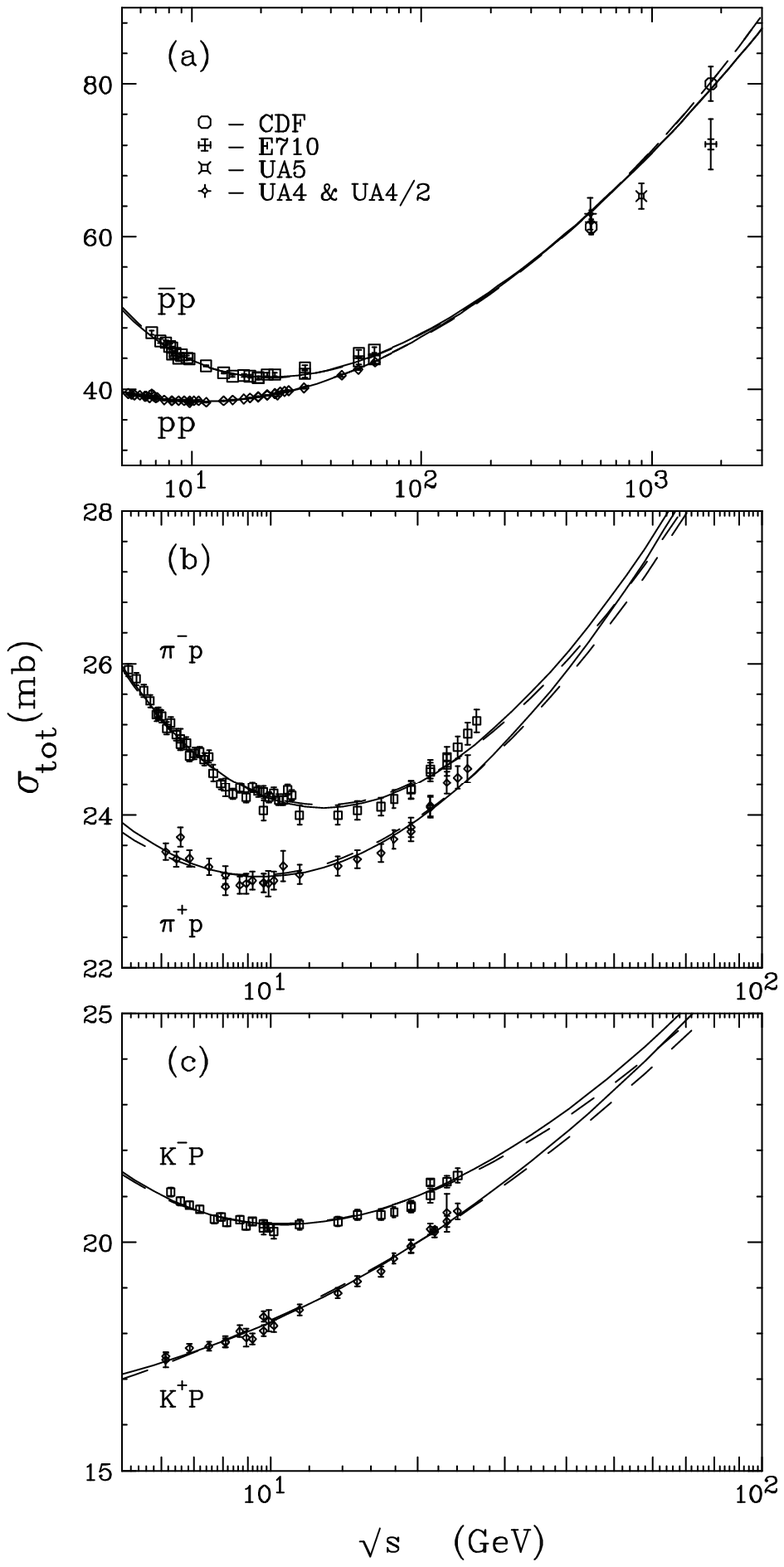,width=0.6\textwidth}\hspace*{-13em}\psfig{figure=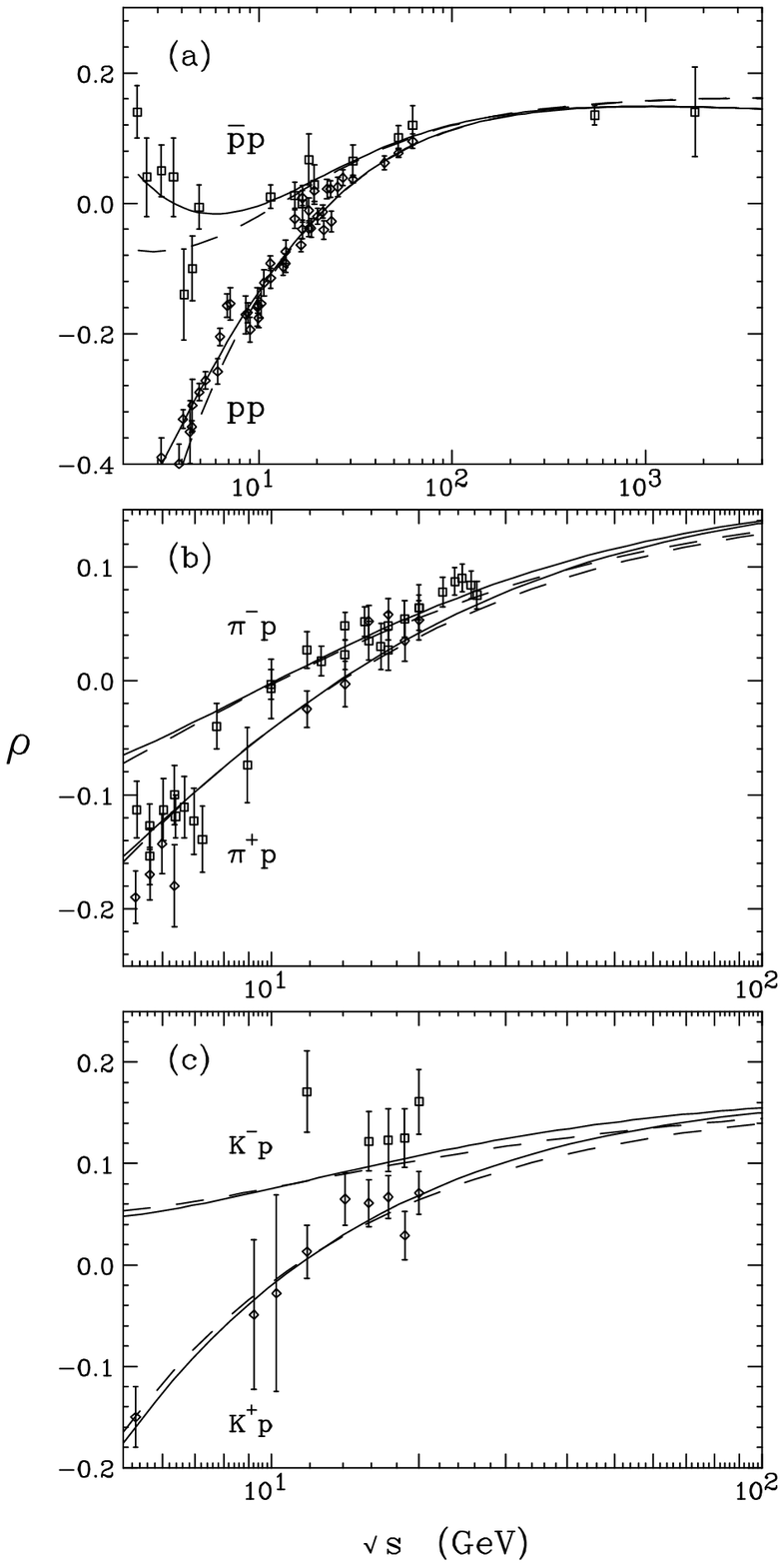,width=0.6\textwidth}\hspace*{-9em}{\vspace*{5em}\psfig{figure=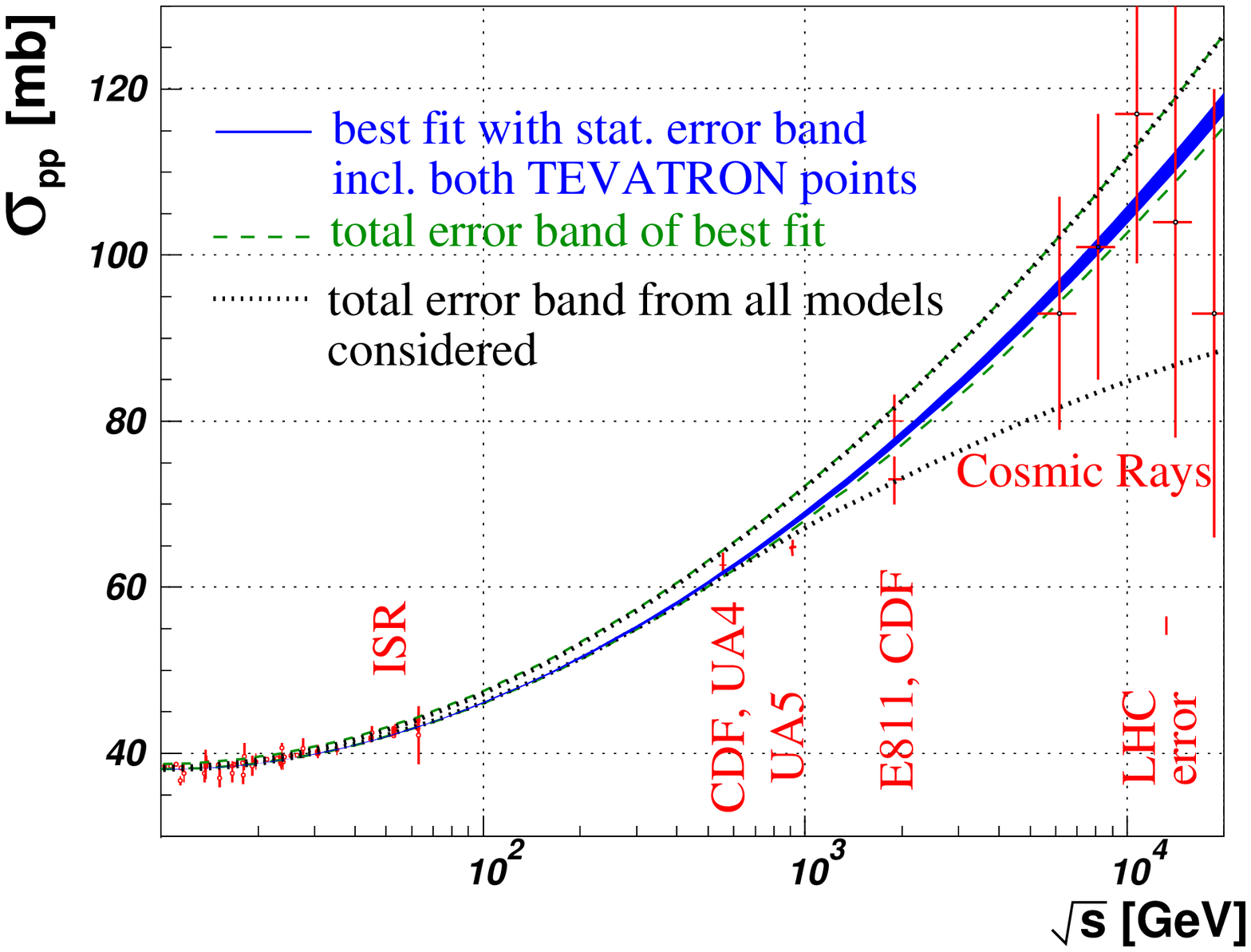,width=0.4\textwidth}}}
\vspace*{1em}
\begin{figure}[h]
\caption{{\em (left / middle)} Simultaneous fit to $p\bar p$, $\pi^\pm$, and $K^\pm$ total cross section and $\rho$-value data using eikonalized (solid) and born level (dashed) amplitudes - the rise of the $p\bar p$ cross section with $\sqrt s$ is ``pulled'' by the rise of the $\pi^\pm$ cross sections and would pass through the CDF point at $\sqrt s=1800$ GeV even if this point were not used in the fit~\cite{ref:CMG}; {\em (right)} fit to $\bar pp$ and $pp$ data with statistical(solid) and total (dashed) error bars - the error band is the 1~$\sigma$ uncertainty due to all parameterizations considered~\cite{ref:Diele,ref:COMPETE}.}
\label{fig:sigmatot}
\end{figure}
\vspace*{-1em}
\paragraph {Measurement issues} Cross sections at the Tevatron have been measured using the luminosity independent method, which is based on simultaneously measuring the total interaction rate, which depends on $\sigma_T$, and the elastic scattering differential rate at $t=0$, which depends on $\sigma_T^2$ (optical theorem):
\begin{equation} 
\sigma_T \propto \frac{1}{L} \left(N_{el}+N_{inel}\right)\;\;\;\;\;\&\;\;\;\;\; \sigma_T^2\sim\frac{1}{1+\rho^2}\frac{dN_{el}}{dt}|_{t=0}\;\;\;\Rightarrow\;\;\;\; 
\sigma_T=\frac{16\pi}{1+\rho^2}\frac{1}{N_{el}+N_{inel}} \frac{dN_{el}}{dt} |_{t=0}
\label{eq:sigma}
\end{equation}
It is evident from Eq.~(\ref{eq:sigma}) that the measurement of $\sigma_T$ crucially depends on the determination of the total inelastic rate, which is hampered by acceptance and background systematic effects. Paradoxically, overestimating the total rate, as for example due to background contamination, leads to smaller elastic and total cross sections, while larger cross sections are obtained by loss of inelastic events. The cross sections  measured at the Tevatron by the CDF and E710~/~E811 collaborations, using different detectors and background evaluation techniques, disagree by more than two standard deviations, most likely due to systematic effects in the measurement of the total rate. Fits of the data using different models also vary, as seen in the examples presented in Fig.~\ref{fig:sigmatot}.  \vglue -1em
\paragraph{Prospects} Elastic and total cross sections are currently being measured at 1.96 TeV by the D0 collaboration at the Tevatron, and plans by the TOTEM collaboration for measurements at the LHC at energies up to 14 TeV are well under way~\cite{ref:TOTEM}. 

\section{Diffraction}
Diffractive events are characterized by the 
presence of at least one large rapidity gap, defined as a  region 
of pseudorapidity
devoid of particles.
A diffractive rapidity gap is presumed to be formed by the exchange 
of a {\em Pomeron} , which in QCD is a  color singlet 
quark/gluon object with vacuum quantum numbers. 
Diffraction in which there is a high momentum transfer 
partonic scattering in the event  
is referred to as {\em hard diffraction}. The goal of diffractive studies is to characterize the QCD nature of the Pomeron. The CDF collaboration has been carrying out a comprehensive program on diffraction since the beginning of Tevatron operations. In this section, we briefly summarize the CDF results, concentrating on aspects that shed light on the QCD nature of diffraction.   

The diffractive processes studied by CDF in Tevatron Run~I (1992-1996) are schematically shown in Fig.~\ref{fig:diagrams}b. A discussion of the results obtained and of their significance in deciphering the QCD nature of the diffractive exchange can be found in~\cite{ref:lathuile}. The most interesting discoveries from the Run~I diffractive program are the breakdown of factorization and its restoration in events with multiple rapidity gaps.    
\vspace*{-1em}
\paragraph{Breakdown of factorization}
At $\sqrt s=$1800 GeV, the SD/ND ratios (gap fractions)   
for di-jet, $W$, $b$-quark, and $J/\psi$ production, as well the ratio of
DD/ND di-jet production, are all $\approx 1\%$.
These ratios are suppressed by a factor of $\sim$10 
relative to predictions based on 
diffractive parton densities measured from DDIS at HERA, indicating a breakdown of QCD 
factorization. There are, however, two interesting features characterizing the data: first, despite the overall suppression in normalization, 
factorization approximately holds among different diffractive 
processes at fixed $\sqrt s\,$, and second,
the magnitude of the suppression is comparable to that
observed in soft diffraction processes relative to Regge theory expectations.
These features indicate that the suppression is not on the hard scale sector but has to do with the formation of the 
rapidity gap. A good description of this feature of the data is provided by the generalized gap renormalization model, which is reviewed in Ref.~\cite{ref:lathuile}. 
\vspace*{-1em}
\paragraph{Restoration of factorization in multi-gap diffraction}
Another interesting aspect of the Run~I results is that 
ratios of two-gap to one-gap cross sections 
for both soft and hard processes appear to obey factorization. This feature 
of the data provides both the clue to understanding diffraction in terms of a composite Pomeron and 
a tool for diffractive studies using processes with 
multiple rapidity gaps, as discussed in~\cite{ref:lathuile}.
\vspace*{-1em}
\paragraph{The Run~II diffractive program} In Run~II, CDF  
is aiming at further deciphering the QCD nature 
of the Pomeron by measuring the dependence of the diffractive 
structure function on $Q^2$, $x_{Bj}$, $t$, 
and $\xi$ (fractional momentum loss of the diffracted nucleon) 
for different diffractive production processes. In addition, 
the possibility of a composite Pomeron is being investigated by studies of 
very forward jets with a rapidity gap between jets.
Another goal of the program is to measure exclusive production rates 
(di-jet, $\chi_c^0$, $\gamma\gamma$), 
which could be used to establish 
benchmark calibrations for 
exclusive Higgs production at LHC.  
Preliminary results from data collected at $\sqrt s=1960$ GeV  
confirm the Run~I results on the diffractive structure function (DSF).
New results from Run~II are the measurement of the 
$Q^2$ and $t$ dependence of the DSF obtained 
from di-jet production, and the measurement of exclusive di-jet and di-photon production rates. 

\begin{figure}[hp]
\hspace*{4em}\hbox{\psfig{figure=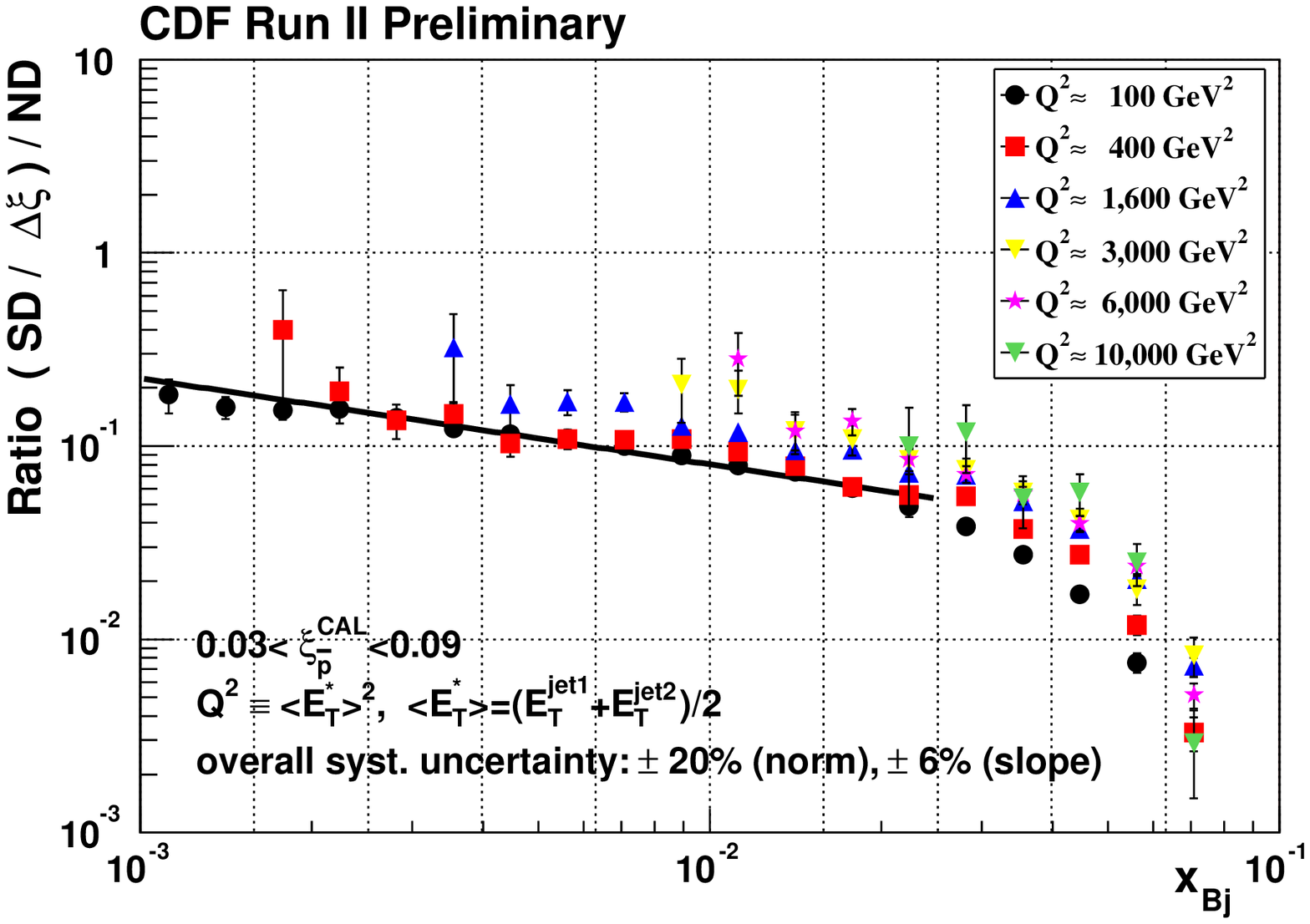,width=0.53\textwidth}\psfig{figure=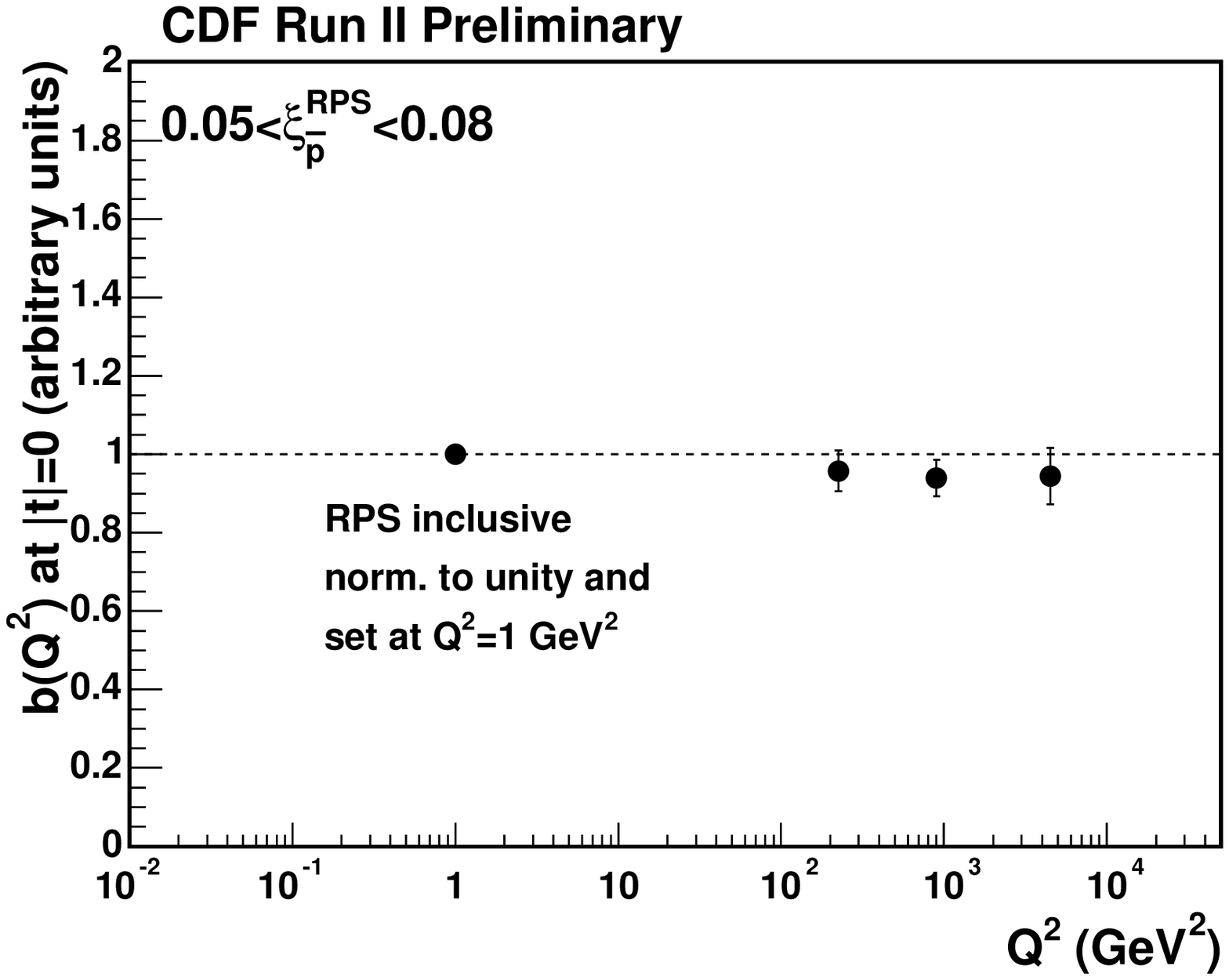,width=0.5\textwidth}}
\hspace*{2em}\caption{
{\em (left)} Ratio of diffractive to non-diffractive di-jet event rates as a function of $x_{Bj}$ (momentum fraction of parton in the antiproton) for different values of $E_T^2=Q^2$;
{\em (right)}b1 slope vs $Q^2$ (normalized to RP). 
}
\label{fig:xbjQ2}
\end{figure}

\newpage
\section{Exclusive Production}
The search for Higgs bosons ranks high in the research agenda of high energy physicists. While the main effort is directed toward searches for inclusively produced Higgs bosons, an interest has developed toward  exclusive Higgs production, $\bar p/p+p\rightarrow \bar p/p+H+p$, which presents several advantages: it can provide clean events in an environment of suppressed QCD background, in which the Higgs mass can accurately be measured by the missing mass technique if the events are tagged by detecting and measuring the momentum of the outgoing proton and (anti)proton. 
However, exclusive production is hampered by expected low production rates~\cite{ref:KMR}. As rate calculations are model dependent and generally involve non-perturbative suppression factor(s), they must be calibrated against processes involving the same suppression factors(s) but have much higher rate that makes them observable at present collision energies and luminosities. Two such  ``standard candle'' processes for calibrating Higgs production rates are exclusive di-jet and exclusive $\gamma\gamma$ production. Both processes are being  studied at the Tevatron, and CDF has reported preliminary cross section results . 
\vglue -0.5cm
\paragraph {Exclusive di-jet production}
The CDF search for exclusive di-jet production is based on measuring the 
di-jet mass fraction, $R_{jj}$, 
defined as the mass of the two leading jets in an event, $M_{jj}$,  
divided by the total mass 
reconstructed from the energy deposited in all calorimeters towers, $M_X$.
The signal from exclusive di-jets
is expected to appear at high values of $R_{jj}$, smeared by resolution and radiation effects.  Events from 
inclusive DPE production, 
$\bar p p\rightarrow \bar p+gap+jj+X+gap$, are expected to contribute to the 
entire $M_{jj}$ region, and any such events within the exclusive $M_{jj}$ range would contribute to background. Two methods are used to extract the signal from the inclusive $R_{jj}$ distribution. In the first method, the background is estimated by Monte Carlo generated events run through detector simulation, while in the second, a distribution from $b$-tagged di-jet events is used to provide the background shape, since exclusive high $E_T$ quark jet production through $gg\rightarrow \bar qq$  is suppressed in LO QCD by the $J_z=0$ selection rule as $m_q/M_{jet}\rightarrow 0$.

\begin{figure}[hp]
\hspace*{1em}\hbox{\psfig{figure=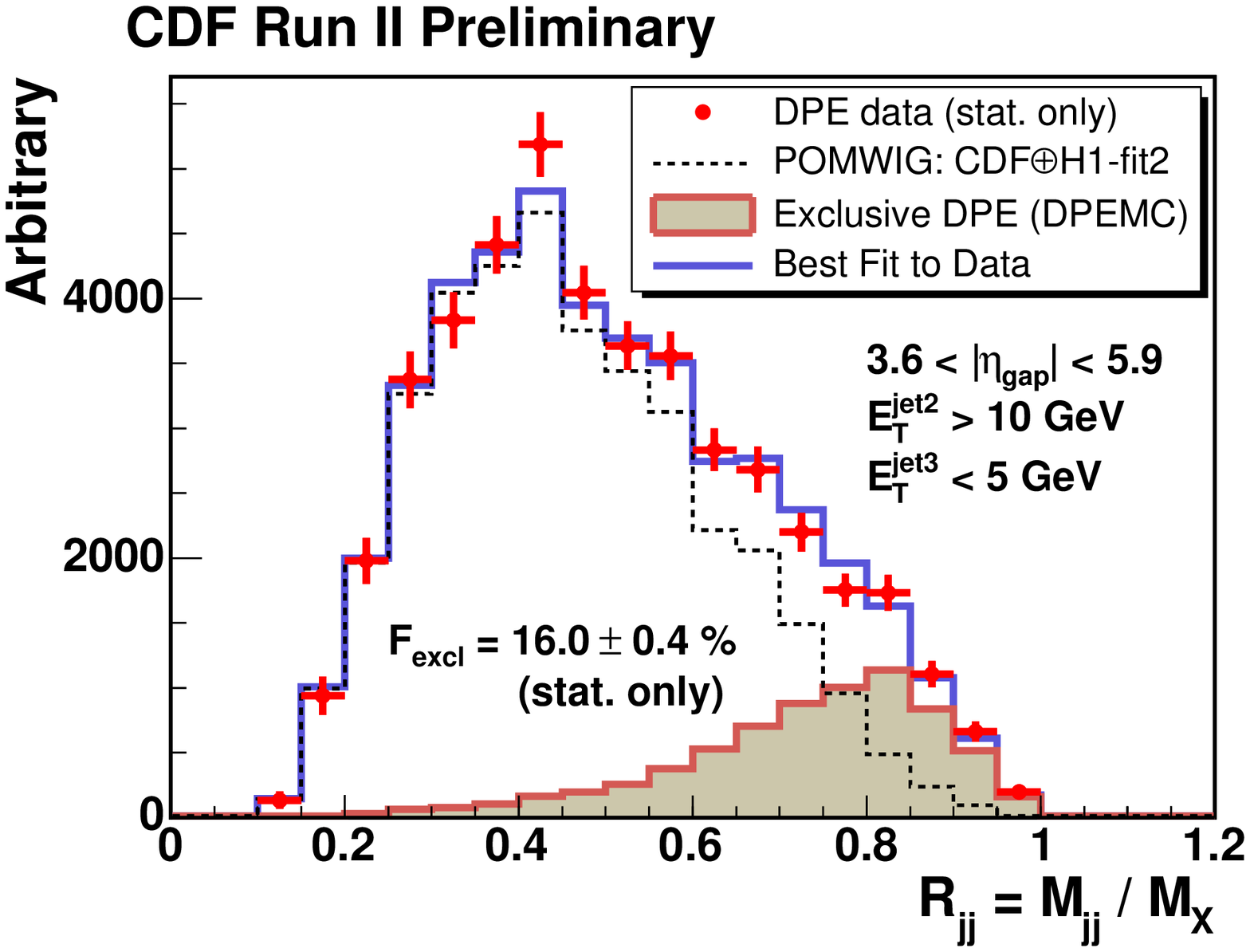,width=0.34\textwidth}\psfig{figure=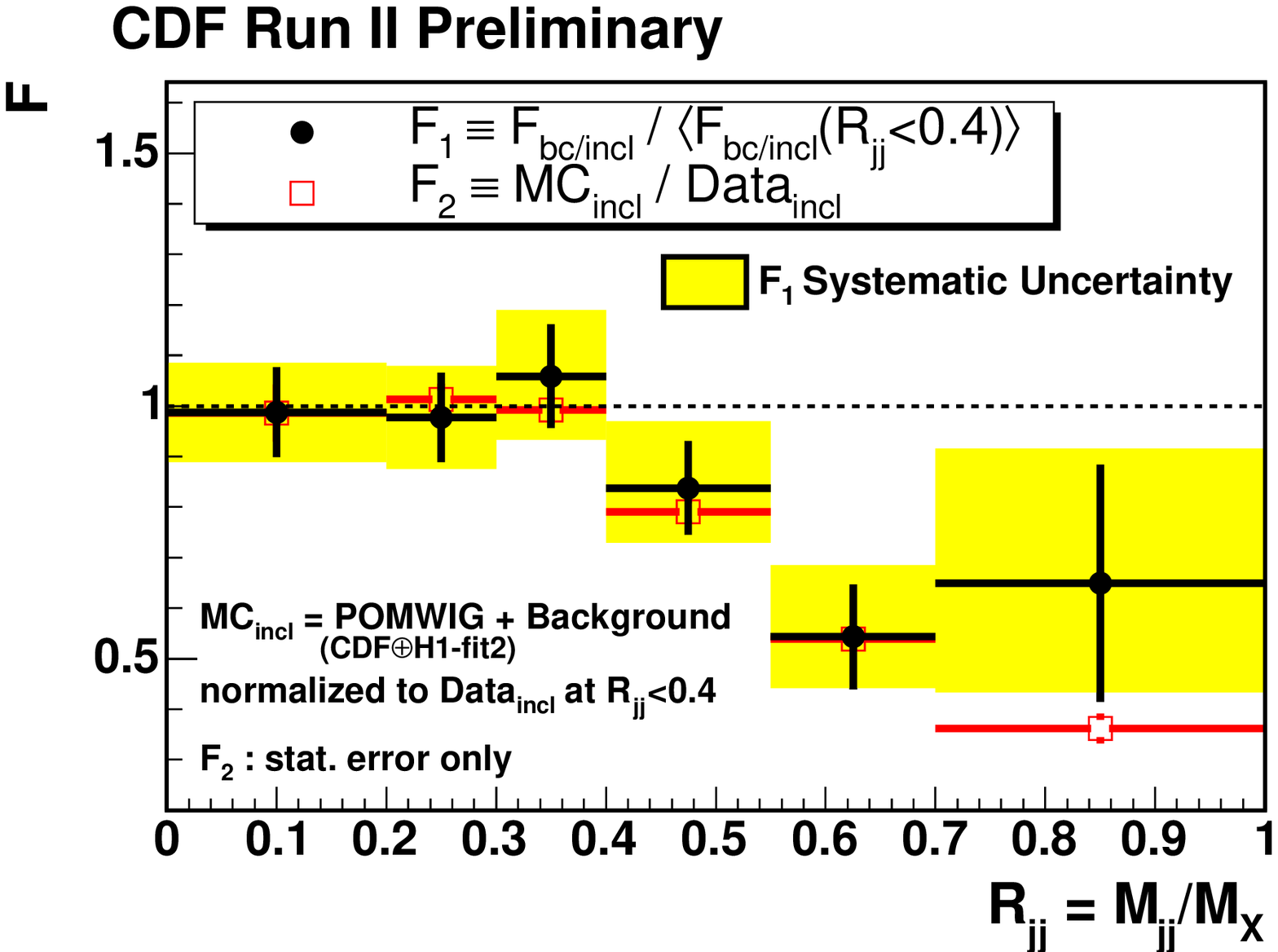,width=0.34\textwidth}\psfig{figure=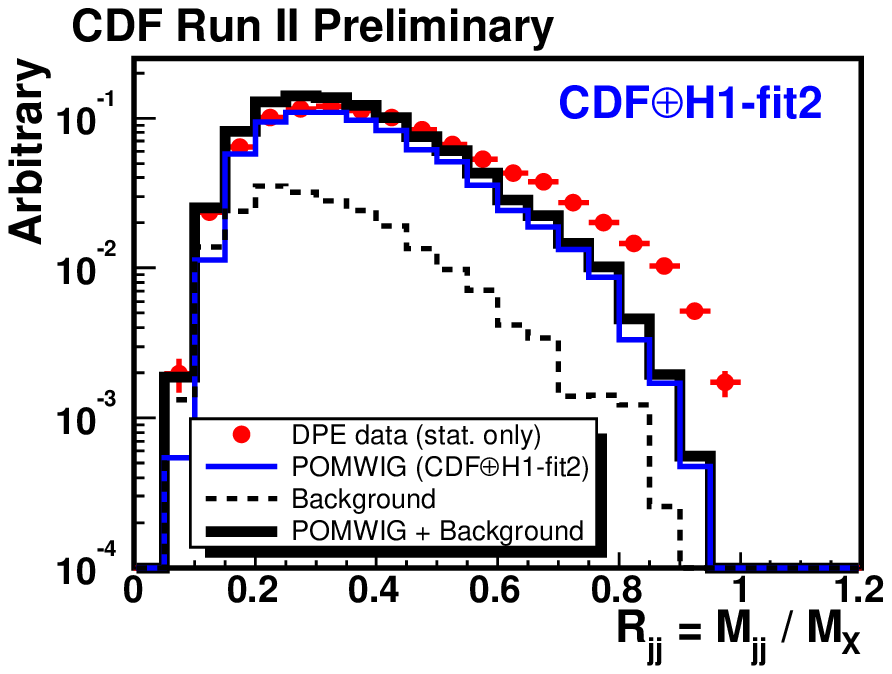,width=0.34\textwidth}}
\caption{
{\em (left)} Di-jet mass fraction in DPE data (points) and best fit (solid) obtained from POMWIG MC events (dashed) 
and exclusive di-jet MC events (shaded);
{\em (center)} normalized ratio of heavy flavor jets to all jets as a function of di-jet mass fraction;
{\em (right)} $R_{jj}$ distribution for data (points) and POMWIG MC prediction (thick histogram), 
composed of DPE di-jet events (thin) and non-DPE events (dashed). Data and MC events are normalized to the same area.
}
\label{fig:excl_JJ}
\end{figure}

Figure~\ref{fig:excl_JJ} shows distributions from the data analysis used to extract the exclusive signal~\cite{ref:michgallDIS}. On the left, di-jet data with DPE topology are fitted with a mixture of inclusive and exclusive MC generated events; in the middle, a suppression at high $M_{jj}$ for $b$-jet events is observed; and on the right, the estimated DPE and non-DPE background contributions are compared with the data. A clear signal is observed above background, which agrees in shape with expectations~\cite{ref:proof}.

\vglue -0.5cm
\paragraph{Exclusive $\gamma\gamma$ production} 
The CDF search for exclusive $\gamma\gamma$ production was performed on a sample of events collected by requiring a high $E_T$ electromagnetic shower in combination with a loose forward rapidity gap requirement. In the data analysis, the rapidity gap requirement was tightened, and a search was performed for events with two high $E_T$ photon showers satisfying ``exclusivity'' requirements~\cite{ref:hamilton}. Three exclusive $\gamma\gamma$ candidate events with $E^\gamma_T>5$~GeV were found with no tracks pointing to the electromagnetic clusters. As a check of the robustness of the rapidity gap requirement, CDF measured the cross section for the  
purely QED process $\bar p+p\rightarrow \bar p+e^+e^-+p$, whose cross section can be reliably calculated. Twelve exclusive $e^+e^-$ candidate events were found in the data with an estimated background of $2.1^{+0.7}_{-0.3}$, yielding 
$\sigma (e^+ e^-)= 1.6^{+0.5}_{-0.3} {\rm (stat)} \pm 0.3 {\rm (syst)}$~pb, which agrees with the an expectation of $1.711\pm 0.008$~pb. For $\gamma\gamma$ production, the 3 candidate events found are to be compared with an expectation of $1^{+3}_{-1}$ based on Ref.~\cite{ref:KMR}.


\section{Diffraction at the LHC}
From a  physics perspective, the LHC will provide a suitable environment for two types of diffraction studies: \\(i) ``standard'' diffraction, with a goal to further probe the QCD nature of the Pomeron,  and (ii) exclusive production, aiming at discoveries of states with vacuum quantum numbers, such as Higgs bosons, in a low background environment. 
\vspace*{-0.25cm}
\paragraph{Standard diffraction}   
The rapidity span at the LHC operating at $\sqrt s=14$ TeV is $\Delta\eta=19$, compared to $\Delta\eta=15$ at the Tevatron. The larger rapidity region can be exploited in designing a program for studies of diffraction based on measurements of multi-gap to single-gap ratios, which are not suppressed by non-perturbative rapidity gap formation effects. 

The following program has been suggested~\cite{ref:dis05LHC}:
\begin{itemize}
\item Trigger on two forward rapidity gaps  of $\Delta\eta_F\geq 2$ 
(one on each side of the 
interaction point), or equivalently on 
forward protons of fractional longitudinal momentum loss 
$\xi=\Delta p_L/p_L\leq 0.1$, and explore the central rapidity region of 
$|\Delta\eta|\leq 7.5$, which has the same width 
as the entire rapidity region of the Tevatron.
In such an environment, the ratio of the rate of 
di-jet events with a gap between jets to that without a gap,  
$gap$+[jet-gap-jet]+$gap$ to $gap$+[jet-jet]+$gap$, should rise 
from its value of $\sim 1$\% at the Tevatron to $\sim 5$\%. 
\item Trigger on one forward gap of $\Delta\eta_F\geq 2$ or on a proton of $\xi<0.1$,
in which case the rapidity gap available for non-suppressed diffractive 
studies rises to 17 units.
\end{itemize}

\paragraph{Exclusive production}
There has been wide interest recently in exclusive production, propelled by the possibility of observing Higgs bosons in an environment of low QCD background, in which the mass of the Higgs boson can be accurately measured from the momentum loss of the scattered forward protons using the missing mass technique~\cite{ref:DeRock}. An international collaboration has been formed to study the feasibility of installing Poman pot detectors in straight sections of the LHC collider located at distances of 420~m from the CMS and/or ATLAS main detectors on both sides of the interaction region. Known as the FP420 (Forward Protons \@ 420 m) project, an R\&D program is under way, aiming at installing detectors for data taking during high luminosity LHC running~\cite{ref:FP420}.
  



\end{document}